\documentstyle[11pt,aaspp4]{article}
\include{epsf}

\makeatletter 
\let\@internalcite\cite
\def\cite{\@ifstar{\citeyear}{\citefull}}
\def\citefull{\def\astroncite##1##2{##1 ##2}\@internalcite}
\def\citeyear{\def\astroncite##1##2{##2}\@internalcite}
\def\citeau{\def\astroncite##1##2{##1}\@internalcite}
\def\citen{\def\astroncite##1##2{##1 (##2)}\@internalcite}
\def\possesivcite{\def\astroncite##1##2{##1's (##2)}\@internalcite}
\def\@citex[#1]#2{\if@filesw\immediate\write\@auxout{\string\citation{#2}}\fi
  \def\@citea{}\@cite{\@for\@citeb:=#2\do
    {\@citea\def\@citea{; }\@ifundefined
       {b@\@citeb}{{\bf ?}\@warning
       {Citation `\@citeb' on page \thepage \space undefined}}%
{\csname b@\@citeb\endcsname}}}{#1}}
\def\@cite#1#2{#1\if@tempswa , #2\fi}
\def\@biblabel#1{}
\makeatother
\newcommand{\ltap}{\hbox{\raise.5ex\hbox{$<$}
\kern-1.1em\lower.5ex\hbox{$\sim$}}}
\newcommand{\gtap}{\hbox{\raise.5ex\hbox{$>$}
\kern-1.1em\lower.5ex\hbox{$\sim$}}}

\begin{document}

\title{{\Huge Evolution of low-mass metal-free stars \\ \vspace{0.2cm}  including effects of
diffusion \\ \vspace{0.5cm} and external pollution}} \vspace{1.5cm}

\author{{\Large A.~Weiss\altaffilmark{1}, S.~Cassisi\altaffilmark{2,1},  
H.~Schlattl\altaffilmark{1}, M.~Salaris\altaffilmark{3,1}}}

\altaffiltext{1}{Max-Planck-Institut f\"ur Astrophysik, Karl-Schwarzschild-Strasse 1, D-85740 
Garching, Germany}
\altaffiltext{2}{Osservatorio Astronomico di Collurania, Via M. Maggini, I-64100 Teramo, 
Italy}
\altaffiltext{3}{Astrophysics Research Institute, Liverpool John Moores University, 
Twelve Quays House, Egerton Wharf, Birkenhead CH41 1LD, UK}

\vspace{1.5cm}

\begin{abstract} We investigate the evolution of low-mass metal-free
Population~III stars. Emphasis is laid upon the question of
internal and external sources for CNO-elements, which -- if present in
sufficient amounts in the hydrogen-burning regions -- lead to a strong
modification of the stars' evolutionary behavior. For the production
of carbon due to nuclear processes inside the stars, we use an
extended nuclear network, demonstrating that hot $pp$-chains do not
suffice to produce enough carbon or are less effective than the
$3\alpha$-process. As an external source of CNO-elements we test the
efficiency of pollution by a nearby massive star combined with
particle diffusion. For all cases investigated, the additional metals
fail to reach nuclear burning regions before deep convection on the
Red Giant Branch obliterates the previous evolution. 
The surface abundance history of the polluted Pop~III stars is 
presented. The possibilities to discriminate between a Pop~II and a polluted
Pop~III field star are also discussed.

\end{abstract}


\clearpage

\section{Introduction}

Little is known about the first generation of stars, which formed out
of primordial material with a composition resulting from the
Big Bang nucleosynthesis.  In particular, the formation
process of Pop~III stars is not understood and important questions
remain unanswered. What was the Pop~III inital mass function?  
Did the low- or
high-mass stars form first?  Did the first Pop~III Supernovae pollute
the still fully convective low-mass pre-main sequence stars?  On the
observational side, no $Z=0$ star has been observed, a fact which is
not surprising given the interaction of potential metal-free stars
with the galactic interstellar medium over more than 10 billion years.
However, there is an increasing wealth of data concerning the ultra
metal-poor halo stars (${\rm [Fe/H]} \ltap -2.5$), which are
definitely metal-poorer than Pop~II stars, whose metallicities cluster
at ${\rm [Fe/H]} \gtap -2.5$.  The metal distribution in the
atmospheres of ultra metal-poor halo stars (UMPHS) displays an almost
pure SNe~II
signature with $r$-process element abundances very similar to the
solar ones (\cite{rnb:96}; \cite{scb:98}).  Apparently, the envelopes
of these stars contain matter processed in only one generation of
massive stars.  Indeed, \citen{shts:98} argue that individual
UMPHS have metal compositions which can be traced back to {\em
individual} Pop~III supernovae of type II. From these results one can
conclude that the present UMPHS, which evidently must be the low-mass
counterpart of a very early generation of stars, might have formed
immediately {\em after} the first SNe.  In this case, they evolved as
extremely metal-poor stars, such as investigated by \citen{cc:93}.
Alternatively, they might be {\em true Pop~III} stars whose envelopes
have been polluted by SNe ejecta after they already had reached the
zero-age main sequence, and therefore had no longer been fully
convective. The internal, nuclear evolution in this case is that of
Pop~III stars, even if the outer parts of the envelope shows the
presence of metals. It is the second case -- polluted low-mass
stars of initial metallicity $Z=0$ (ignoring the $10^{-10}$ level of
initial BBN $^7$Li and $^7$Be) -- which we are investigating in this
paper.  Our aim is to provide a theoretical background to assess the
evolutionary state of the UMPHS.

There is a significant difference from the structural point of view
between extremely metal-poor stars
and those of zero metallicity, as has been shown in all existing
papers dealing with the evolution of metal-free stars (\cite{dant:82};
\cite{gd:83}; \cite{eek:85}; \cite{fih:90}).  The reason is 
the absence of CNO-elements. While in low-mass Pop~II stars 
during the final stages of the main-sequence (MS) evolution the
CNO-cycle is stabilizing the 
core at moderate temperatures, the sole operation of the
$pp$-chains in Pop~III stars leads to significantly hotter cores
toward the end of core hydrogen burning.  (For more massive Pop~III stars,
see \cite{ec:71}; \cite{efo:83}; \cite{arn:96}. Recall that for
non-zero metallicity, the CNO-cycle is not only contributing but
dominating the energy generation.) 
Similarly, the hydrogen burning shell of the post-MS stars
shows much higher temperatures as for the Pop~II case where the
CNO-cycle provides the dominating energy source.  This implies that if
the star had any source supplying enough CNO, it can convert into
an extremely metal-poor star and might evolve -- after a possible
transition phase -- in a more standard fashion. The onset of the
CNO-cycle is a rather drastic event leading to excursions in the
Hertzsprung--Russell--diagram and to transient convective regions with
mixing episodes. The critical mass
fraction has been shown to be $\approx 10^{-10}$ and results from the
fact that at $T \approx 10^8$~K about $ 10^{-8}$ of the solar
CNO-abundance is sufficient to lead to the same energy generation as
in a star of solar metallicity.

As it has been demonstrated in the early works on Pop~III stars, 
the temperatures at
the end of core hydrogen burning in a $1\,M_\odot$ star (\cite{dant:82};
\cite{fih:90}) are high enough to allow the production of the critical carbon 
abundance 
through $3\alpha$-reactions. Results at lower masses are inconclusive (\cite{dant:82};
\cite{gd:83}). 
In \S~2 we will investigate in detail whether other nuclear chains could be
additional sources for carbon production, most notably those that start from
$pp$-chain elements like $^7$Li.  This is to ensure that the 
{\em in situ} production of carbon is treated correctly in our stellar evolution
calculations, because it is crucial for predicting whether and when Pop~III 
stars can convert themselves into extreme Pop~II stars. In \S~3 we will 
present such up-to-date calculations of Pop~III models, which
are partially a repetition of the classical work cited, but are also an
extension to a variety of masses.  In particular we will discuss for which
stellar mass {\em in situ} production of carbon is possible at all.

As the second source for CNO-elements we investigate that resulting
from the assumed pollution by nearby supernovae.  While the pollution
initially affects only the outermost (possibly convective) envelope,
atomic diffusion, which is known to be a significant physical
process in the Sun, could transport the added metals to the
hydrogen-burning regions. This external source of carbon could in
particular affect the hydrogen burning shell in evolved stars. \S~4 we
will present calculations of Pop~III stars evolution under these
assumptions and discuss our results in
connection with the observations of UMPHS. A section devoted to the conclusions 
follows.

\section{Carbon production by nuclear reactions} 

The dominant reaction for primary carbon production in stars is the
$3\alpha$-process. Hydrostatic helium burning is usually taking place
at $T\approx 10^8$~K, where enough energy is released to provide
stellar luminosities. However, already at lower temperatures
some $3\alpha$-reactions occur. Since hydrogen burning
temperatures in Pop~I and II stars are below $8\cdot10^7$~K, these
two burning phases can safely be assumed to be well separated.  The
situation is different in Pop~III stars, where the central temperature
$T_c$ is higher and already before the exhaustion of protons can
approach values of $7\cdot 10^7$~K or above (see \S~3 and
\cite{fih:90}). 

At this temperature and at the typical densities of late main-sequence
phase cores ($10^5\,{\rm g}\,{\rm cm}^{-3}$) the lifetime of helium
against $\alpha$-capture is of order $10^{12}$~yrs.
Over the remaining $10^8$ years of 
core hydrogen burning the critical amount of carbon of
$10^{-10}$ (mass fraction) can easily be produced, therefore.
For standard Pop~II stars this might happen as well, although in
smaller amounts, but the additional primary carbon produced is
negligible compared to the one already present in the initial
mixture. 

The $3\alpha$-process is not the only way to produce carbon at high 
temperatures. \citen{mit:85} investigated ``unconventional $^{12}$C production 
in Pop~III stars'' via $\alpha$-captures on the light elements $^7$Be and $^8$B, 
which are present in equilibrium abundances in the pp-II and pp-III chains. The 
results of those $\alpha$-captures would be $^{11}$B or $^{11}$C, which under 
subsequent $p$-captures (note that this limits the ability of 
these reaction chains to produce carbon to stages
before the end of hydrogen burning) would create $^{12}$C. Since these reactions 
dominate the $3\alpha$ process at $T\ltap 7.6\cdot 10^7$~K, they might be able 
to create carbon in sufficient amounts even before the latter process,
or -- in stars of lower mass not reaching the critical temperatures
for the $3\alpha$  process -- might be the only path to carbon
production before the end of core hydrogen burning.

Alternatively, in hot pp-chains, $^8$B does not $\beta$-decay but reacts via 
$^8{\rm B}(p,\gamma)^{9}{\rm C}$. By further $\alpha$-capture $^{13}$N can be 
created. \citen{wgg:89} have investigated these hot pp-chains in detail 
with a nuclear reaction network. Besides the classical pp-I, -II, and -III 
chains they identified -- as function of $T$ and $\rho$ -- additional chains 
dominating for hotter conditions. The sequence of reactions proposed by Mitalas 
(1989) was included (termed rap-II and rap-III). Wiescher et al.\ (1989) were 
interested in whether the hot pp-chains could produce CNO-isotopes sufficiently 
fast such that Pop~III supermassive  stars can experience a thermonuclear 
explosion during the fast core collapse. For this specific question they found 
the rap-processes to be important only for temperatures in excess of $10^8$~K 
(see their Fig.~7).

\begin{figure}[ht]
\centerline{\epsfxsize=0.75\hsize\epsffile{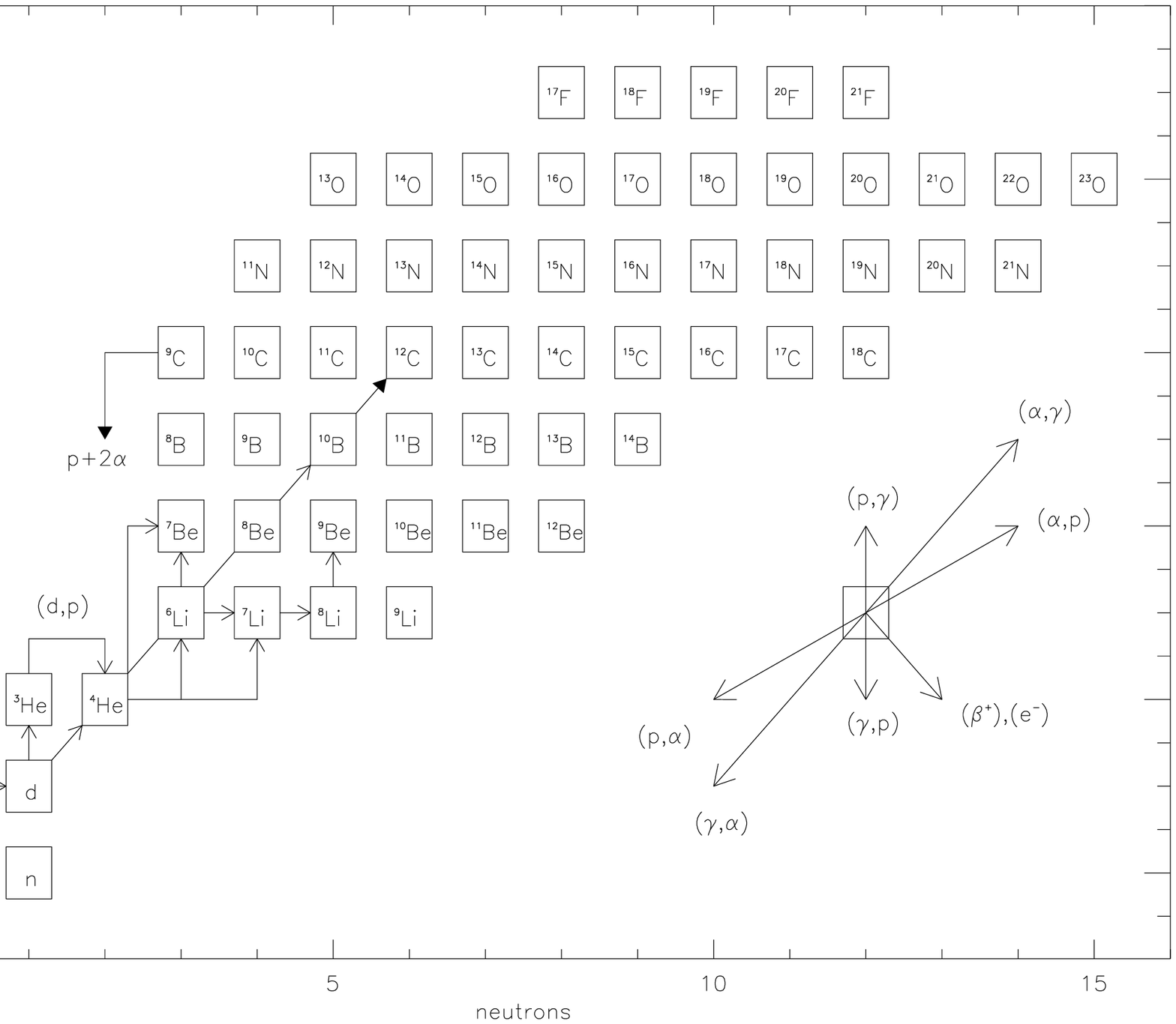}}
\caption{The network used for investigating contributions to carbon
production (see text for explanations)}
\protect\label{f:netstruc}
\end{figure}

In the present case, however, the problem is related to the
competition between pp-chains and $3\alpha$ reactions in producing
carbon during the final phases of core hydrogen burning, i.e.\ on
nuclear time-scales. In order to answer this question and to determine
which reaction chains have to be included in the network part of the
stellar evolution code, we used
the one-zone nuclear reaction network of \citen{wt:90} and included
all potentially relevant reactions\footnote{We will use the term {\em
full network} to denote the one-zone extended nuclear reaction
network and call the more limited network used in the
stellar evolution code  the {\em stellar evolution network}}. 
The reaction rates were taken from the most recent update of
the reaction rate library of F.-K.~Thielemann (\cite{tat:87};
\cite{fkt:96}).  The full network is shown in
Fig.~\ref{f:netstruc}. In the lower right corner standard reactions
($(p,\gamma)$, $(\alpha,p)$, etc.) except those involving neutrons are
shown. They are included for all nuclei. In addition, several (but not
all) other reactions not fitting into this standard reaction scheme are
indicated. As an example, consider the $3\alpha$-reaction indicated by
a long arrow with solid arrowhead. However, also the
$\alpha(\alpha,\gamma)^{8}{\rm B}$ is included as an individual
reaction (open arrowhead). An important reaction is $^9{\rm C}
\rightarrow p + 2 \alpha + \beta^+$, which inhibits the artificial
buildup of $^9{\rm C}$, from which $^{12}{\rm C}$ would become
possible. Some arrows indicate several possible reaction paths, e.g.\
from $^4{\rm He}$ to $^7{\rm Be}$ we have both an $^3{\rm He}$ capture
or an $(\alpha,n)$-exchange included. In the lower left corner we also
included reactions involving $d$, for example $^3{\rm He}(d,p)^4{\rm
He}$ or $d(d,n)^3{\rm He}$. Inverse reactions are always taken into
account as well. The upper right corner contains nuclides included for
the CNO-cycles.

The full network needs as input $T(t)$ and $\rho(t)$, which we took
from selected evolutionary calculations. Since these were done with
our stellar evolution network including only standard pp-, CNO- and
$3\alpha$-reactions (but treating all of them simultaneously), the
full network might be considered to be inconsistent. However, up to a
relative mass fraction of CNO-elements of $X_{\rm CNO} < 10^{-10}$,
the energy production of the star and therefore its temperature
evolution will not be affected and all other effects resulting from
the production of carbon (e.g.\ changing the molecular weight) are
utterly negligible. In addition, the full network calculations have
only exploratory character to identify those reactions which must be
included in the stellar evolution code for the proper treatment of
carbon production.

\begin{figure}[ht]
\centerline{\epsfxsize=0.9\hsize \epsffile{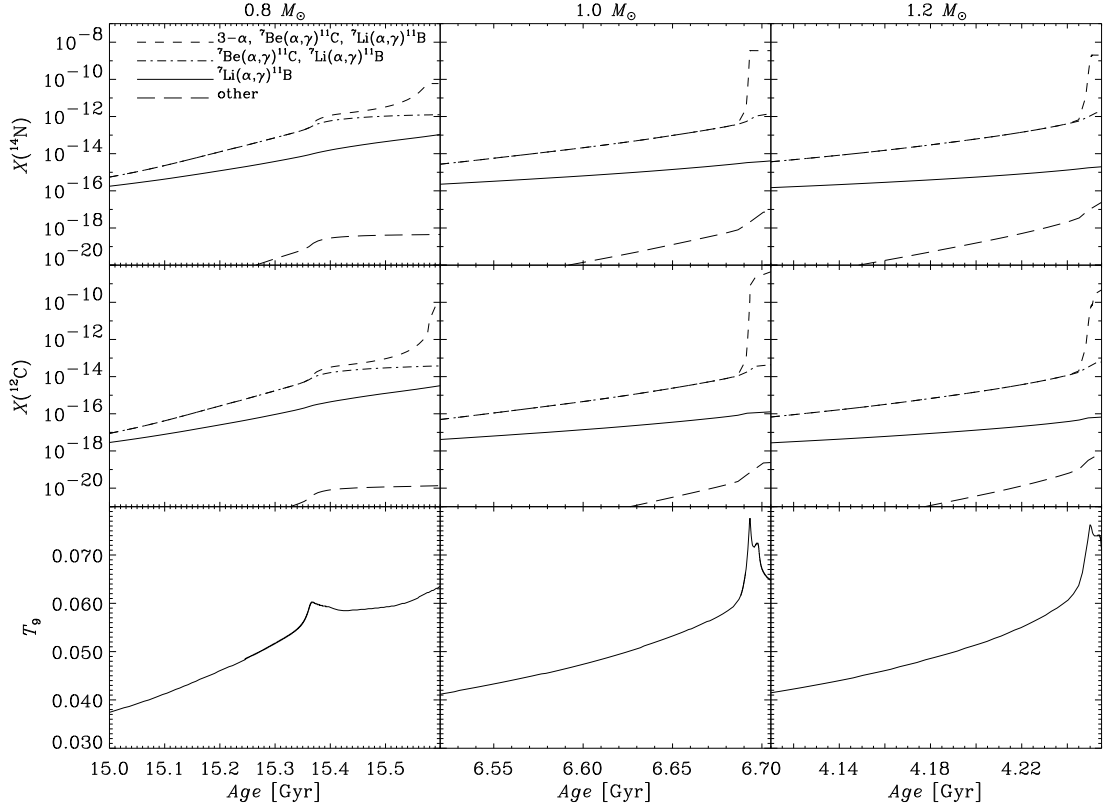}}
\caption{Result of the nuclear network calculations for three selected
stellar masses. Shown is the $^{14}{\rm N}$ (top row) and $^{12}{\rm
C}$ (middle row) mass fraction as well as the central temperature
(bottom row; given as $T_9 \equiv T/10^9$~K) as functions of time}
\protect\label{f:netres}
\end{figure}

Primordial $^7{\rm Li}$ and $^7{\rm Be}$ -- both present at
a level of $10^{-10}$ -- are burnt by
$\alpha$-captures at the beginning of the main-sequence
phase or, equivalently, at the very beginning of the full network
calculations. \citen{mit:85} already demonstrated that their primordial
presence does not lead to carbon production during the
pre-main-sequence phase. Carbon itself is produced in standard BBN
only at a level of $10^{-15}$ or lower; in inhomogeneous BBN this can
rise under very high $n/p$-ratios to $10^{-12}$ at most
(\cite{tso:93}). For the full network we assumed a zero abundance.

The results of the full network calculations are summarized in
Fig.~\ref{f:netres}, which displays the creation of $^{12}{\rm C}$ and
$^{14}{N}$ as a function of time for the last phases of the
main-sequence evolution of 3 stars of masses 0.8, 1.0 and
$1.2\,M_\odot$.  The bottom panel shows the run of temperature as
taken from the stellar evolution calculations including the
temperature rise due to the rapid conversion of carbon to nitrogen. 
The three lines in the upper two panels correspond to the
production if all important processes -- $3\alpha$, $^7{\rm
Be}(\alpha,\gamma)^{11}{\rm C}$ and $^7{\rm
Li}(\alpha,\gamma)^{11}{\rm B}$ -- are included (dashed line), if
$3\alpha$ is excluded (dot-dashed), and if only the last one is
considered (solid).  This latter chain -- $^7{\rm
Li}(\alpha,\gamma)^{11}{\rm B} (p,\gamma) ^{12}{\rm C}$ -- turned out
to be the next one in the order of decreasing carbon production
effectivity.  
For all three masses, the $^7{\rm
Be}(\alpha,\gamma)^{11}{\rm C}$ chain is the only relevant one for the
initially lower temperatures, as is evident from the fact that the
dashed and dot-dashed lines lie on top of each other (i.e.\ the
neglect of the $3\alpha$-process has no influence), while the solid line
($^7{\rm Li}(\alpha,\gamma)^{11}{\rm B}$ only) falls below them,
even if by less than one order of magnitude. At increasing temperatures
towards the end of core hydrogen burning, the $3\alpha$-process
quickly becomes the most important source for carbon, which is
partially processed to nitrogen, in particular in the more massive
stars. The other chains' contributions decline such that the
corresponding lines almost level off. This is due to the exhaustion of
protons needed for carbon production in these chains. Note that at
earlier times, when enough protons are still available, the conversion
of $^{12}{\rm C}$ to $^{14}{\rm N}$ -- the CN-subcycle -- is possible
such that $^{12}{\rm C}/^{14}{\rm N}$ is in equilibrium at $\approx
10^{-2}$. With the exhaustion of protons and the additional $^{12}{\rm
C}$-source through $3\alpha$, carbon finally is exceeding nitrogen in
abundance.

While the $^7{\rm Be}(\alpha,\gamma)^{11}{\rm C}$ is the major path to
$^{12}{\rm C}$ and $^{14}{\rm N}$ for some time, in no case this
process is able to create the critical abundance of $\approx
10^{-10}$. Ignoring all processes except $3\alpha$ therefore
introduces only insignificant errors. The relevant carbon production
will happen (if it does so at all) through $3\alpha$ at the end of
core hydrogen burning and the carbon missing at that time due to the
neglect of the other processes (at a level of $10^{-12}$ or even less)
will quickly be added.  Note that the $0.8\,M_\odot$ star just manages
to reach the critical $^{12}{\rm C}$ abundance. Due to the lower core
temperatures, stars of even lower mass will fail to create enough
carbon before the exhaustion of protons such they will not be able to
initiate the CNO-cycles. Also, their main-sequence lifetimes are
longer than the age of the universe.

{From} the detailed network computations we have performed we therefore conclude 
that indeed it is safe to include only the $3\alpha$ process in the stellar 
evolution code if one accepts errors of 1\% or less in the detailed $^{12}{\rm 
C}$ and $^{14}{\rm N}$ production history (which is well contained in the 
uncertainty of the $3 \alpha$ rate). However, it is also evident that both H- 
and He-burning reactions must be treated simultaneously in the
network.

\section{The evolutionary computations: ``standard'' models}

The evolutionary properties of extremely metal-poor stellar structures
have been the subject of accurate investigations since the early
1970s, thanks to the pioneering work by \citen{ec:71}. The early
investigations have usually been devoted (see the introduction) only
to selected evolutionary masses or quite narrow mass ranges; the first
complete survey of the evolutionary behavior of extremely metal-poor
objects in a quite large range of mass (from low-mass to intermediate
and massive stars) has been carried out by \citen{cc:93}, and was
later completed by \citen{cct:96} by extending the numerical
computations for low-mass stars to more advanced evolutionary stages,
as the central He burning and the double shell H and He burning
phases.  Following \citen{ahs:88} the metallicity employed in the
computations was $Z=10^{-10}$, adopted as an upper limit for the
cosmological production of heavy elements in inhomogeneous Big Bang
nucleosynthesis.  In the latter two works a big effort was devoted to
investigate all the peculiar evolutionary features related to the
paucity of heavy elements in the stellar matter; for this reason a
detailed description of the main evolutionary properties of
Population~III objects will not be repeated here and we refer the
interested reader to the quoted papers and reference therein.

In the present work, we decided to adopt a "true" metal free ($Z=0$)
chemical composition.  As noted above, the only elements present at a
level of $10^{-10}$ in the Pop~III primordial material are $^7{\rm
Li}$ and $^7{\rm Be}$, which, however, are burnt to helium in the
largest parts of the stars already during the pre-main-sequence phase.

The evolutionary models  have been computed by 
adopting an updated version of the FRANEC evolutionary code (\cite{cs:97}; 
\cite{sc:98}).
As for the input physics, OPAL radiative opacity tables (\cite{irw:92};
\cite{ri:92}) were adopted for temperatures higher than 10,000 K, while 
for lower temperatures the molecular opacities provided by 
\citen{af:94} have been used. Both high and low temperature opacity tables assume
a scaled solar heavy elements distribution (\cite{gre:91}) when $Z>$0.
The equation of state has been taken from \citen{s:88},
supplemented by a Saha EOS at lower temperatures.
The outer boundary conditions for the stellar models have been evaluated by assuming the 
$T(\tau)$ relation provided by \citen{kswa:66}. In the superadiabatic 
region of the stellar envelope a mixing length value of 1.6 pressure
scale heights has been adopted. 

We emphasize that for some selected evolutionary sequences the
numerical computations have also been performed with the
Garching evolutionary code (e.g.\ \cite{schw:99}) in order to verify
the reliability of the present results. In all cases we have obtained
good agreement between the two set of "parallel" computations; small
differences are due to some differences in the adopted physical
inputs. For example, the OPAL EOS (\cite{rsi:96}) is being used in the
Garching code.

Following the suggestions by \citen{ec:71} and by \citen{cc:93}, local
equilibrium values have been adopted for the $^{3}{\rm He}$ abundance,
independent of the occurrence of convective mixing. If and when the
CNO-cycle becomes operative, equilibrium abundances among the various
nuclei have been assumed, as derived by solving - for any given
temperature and density - the set of equations describing the
equilibrium among $p$-captures and $\beta$-decay rates.  
In both cases, the approach adopted for the chemical equilibrium treatment
follows from the evidence that - under the peculiar physical
conditions existing in extremely metal-poor stars - the equilibrium
timescales are much shorter than the characteristic mixing times (for
a detailed discussion on this point we refer to \cite{csac:78})

Taking into account the evidence presented in the previous section
that the $3\alpha$-reaction is by far the major contributor to
$^{12}{\rm C}$ formation, the evolutionary computations have been
carried out by adopting a nuclear network accounting for both H and
He-burning, but neglecting any possible unconventional nuclear
reactions branch.  For all the models an initial helium abundance
equal to $Y=0.23$ -- in agreement with the current estimations on the
primordial helium abundance -- has been adopted.

We have considered stellar masses ranging from $0.8M_\odot$ to
$1.2M_\odot$ (in steps of $0.1M_\odot$), and all the evolutionary
sequences have been followed from the initial Zero-Age Main Sequence
(ZAMS) until the He-burning ignition at the tip of the Red Giant
Branch (RGB); in the following, we will describe the properties of
$1M_\odot$ stellar models, taken as being representative of this mass
range. The evolution of the stars in the considered mass range is
quite similar, and the particular choice allows us to make useful
comparisons with the results from previous investigations.

\begin{figure}[ht]
\centerline{\epsfxsize=0.6\hsize \epsffile{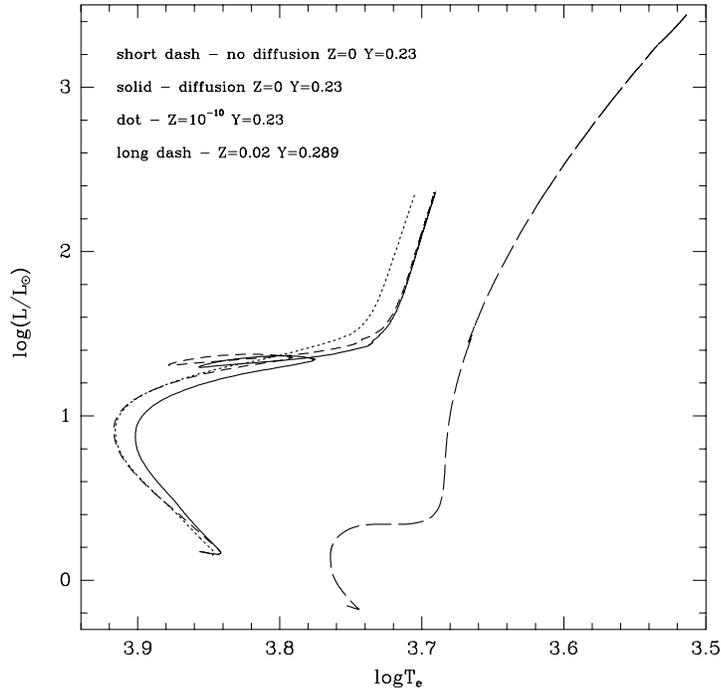}}
\caption[]{The evolutionary paths in the H-R diagram for $1M_\odot$ stellar models,
computed under different assumptions about the initial chemical composition: canonical
metal-free model (short dashed), standard metal-free model accounting for atomic diffusion
(solid), canonical $Z=10^{-10}$ model computed by \citen{cc:93} (dotted) and $Z=0.02$
for comparison (long dashed). All sequences have been
followed until the iginition of helium in the degenerate cores}
\protect\label{f:hrd1}
\end{figure}
For each mass we have computed two different cases without external
pollution: one {\em without atomic diffusion}, which we denote as
'canonical' and one {\em with atomic diffusion} (the diffusion has
been treated according to the formalism by \cite{thoul:94}). The
second case we call `standard' following the contemporary definition
of the physics for the standard solar model.  In addition (see next
section), we have computed some evolutionary sequences taking into
account both atomic diffusion {\em and} external He and heavy elements
pollution.  This case we will call in the following `pollution' or
similar. It is worth noticing that this is the first theoretical
investigation of Pop~III stars accounting in a self-consistent way for
the effect of atomic diffusion alone, and of external pollution plus
atomic diffusion.

The evolutionary paths of both the $1M_\odot$ canonical model and the
one computed accounting for helium diffusion (standard), are plotted in
Fig.~\ref{f:hrd1}. For comparison, we have also plotted the
evolutionary tracks for an extremely metal-poor chemical composition
($Z=10^{-10}$) as given by \citen{cc:93} and for a solar composition,
calculated by us with the same program.

Comparing our canonical model with the one by \citen{cc:93} it is
worth noticing the fine agreement between the two computations, as far
as the MS evolution is concerned: the two tracks are
perfectly overlapping until the early evolution along the sub-giant
branch (SGB). The age of the two models at the Turn Off (TO) is almost
the same: $t_{\rm H}=6.43$ Gyr for the models computed by
\citen{cc:93} and $t_{\rm H}=6.31$ Gyr for our models; the difference
being of the order of a negligible 2\%.  This result is not surprising
since the main difference between the present computations and the
ones performed by \citen{cc:93} is only the use of different opacity
tables. More in detail, \citen{cc:93} employed the \citen{cox:76}
opacities for $T<$12000 K, and those by \citen{laol:77} at higher
temperatures.  The evolutionary timescales on the MS and the MS
location may be affected by the high-temperature opacities employed in
the computations, but the difference between OPAL and \citen{laol:77}
data is negligible at such low metallicities (see also the 
discussion in \cite{cc:93}).  On the other hand, the difference in the
location of the RGB is due to the higher values of the low-temperature
opacities employed in the present calculations, since the low
temperature opacities determine the position of the RGB in the
Hertzsprung-Russell-diagram (HRD) entirely.

Fig.~\ref{f:hrd1} highlights the evolutionary event, peculiar of true
Pop~III stars, which occurs during the transition to the RGB: a flash in the
H-burning region, which leads to a blue loop HRD. The same result has
been obtained also by \citen{fih:90} by computing a $1M_\odot$ metal-free model.
The physical reasons for the occurrence of the flash have been discussed 
by \citen{fih:90} and
can be summarized as follows. Close to the exhaustion of hydrogen at
the center of the star, the 
core strongly contracts, increasing temperature and density, thereby 
counterbalancing the effect of the continuous hydrogen depletion on energy production. 
Due to this increase in both central density and temperature, there is
an increase in the efficiency of the triple-$\alpha$ reactions and, in turn, of the
abundance of carbon nuclei (cf.\ Fig.~\ref{f:netres}). As a
consequence, the energy delivered by the CNO cycle 
strongly increases during this phase. A thermal runaway is produced and a flash occurs
which produces the loop of the stellar track in the HRD. In
this phase, the star develops a convective core. The ensuing
expansion of the inner stellar regions produces a 
significant decrease in both density and temperature, which reduces the
$3\alpha$ nuclear rate and, as a consequence, also the energy produced
by the CN-conversion. When the central abundance of hydrogen drops to zero
the convective core disappears.

Even if the general characteristics of this phenomenon are quite similar to the ones
obtained by \citen{fih:90}, there are also remarkable differences, which are perhaps
due to differences in the physical inputs and in the numerical treatments.
In both models, the flash along the SGB occurs
when the central abundance of hydrogen is of the order of $X_c\approx0.0006$. However,
the maximum size of the convective core during the thermal runaway is about 50\%
lower in our computations than in the model computed by \citen{fih:90}
($\sim0.11M_\odot$ in comparison with $\sim0.2M_\odot$). This means that in
our model the energy flux produced by the flash is significantly lower in comparison
with previous results. Another significant difference between the two sets of
computations is related to the fact that during the runaway the
hydrogen-burning rate via the CN cycle is able to exceed the
contribution provided by the {\sl pp}-chain in the model produced by  
\citen{fih:90}. In our computations we find that
when the energy delivered by the CNO cycle reaches its maximum it is
about a factor 
3 lower than the {\sl pp}-chain contribution. Also the maximum
helium-burning rate in our models ($L_{\rm He}\approx25.7\cdot10^{-8}
L_\odot$) is about two orders of magnitude smaller than the value 
obtained by \citen{fih:90}.

Due to the general properties of this phenomenon, we suggest that it is not really 
resembling a He flash like the one occurring at the RGB tip (as suggested by 
\cite{fih:90}) but that it is more similar to a CNO flash. It is interesting
to notice that the chemical 
abundances of CNO elements are $X_{^{12}{\rm C}}=6.50\cdot10^{-12} -
X_{^{14}{\rm N}}= 2.19\cdot10^{-10} - X_{^{16}{\rm O}}= 
2.77\cdot10^{-12}$ at the maximum flash energy production.

Comparing this result with the models provided by \citen{cc:93},
one notices that the CNO flash along the SGB is missing in the
latter computations. A quick comparison between the two models shows
that both density and temperature values in the inner core of the
stars are significantly lower in the models of \citen{cc:93}.  As we
verified with a specific evolutionary computation, this is a result of
the fact that \citen{cc:93} have assumed a non-zero primordial
abundance of heavy elements: even this extremely low metal abundance
($Z=10^{-10}$) is able to allow the CNO-cycle to operate, such that --
at the exhaustion of the central hydrogen abundance -- the central
temperature remains low enough to avoid the thermal runaway.
The initial chemical abundances of CNO elements in
the \citen{cc:93} model are quite similar to those of our
zero-metallicity model {\em after} the thermal runaway; in this sense, a
metal abundance of the order of $Z=10^{-10}$ seems to be a "critical"
metallicity for getting the thermal runaway along the SGB. 
In any case, one has to bear in mind that the CNO-flash along
the SGB is a secondary evolutionary feature which only slightly
increases the central hydrogen burning phase duration (of order
$10-20\cdot 10^6$~yrs). Therefore, for
general purposes as, for instance, population synthesis, the
theoretical framework developed by \citen{cc:93} and \citen{cct:96}
would be applicable also to canonical $Z=0$ models.

The comparison between the canonical and standard model
accounting for atomic diffusion shows the expected changes, i.e.\ a
moderate decrease of the TO luminosity and of the central hydrogen
burning lifetime, and a slight shift of the MS toward lower effective
temperatures. It is evident that helium diffusion has no effect on
the occurrence of the CNO-flash at the end of the central hydrogen
burning phase.  It is also worth noticing that one has to expect that
the efficiency of atomic diffusion during the MS evolution of low mass
stars increases when decreasing the stellar metallicity, since the
diffusion in the stellar envelopes is larger due to the thinner
convective envelopes.  Therefore, for fixed diffusion coefficients the
effect of atomic diffusion is largest in metal-free stars.  This is a
quite important point to bear in mind when we will discuss the effect
of external pollution on metal-free stars.

As far as the evolution along the RGB is concerned, we did not find
any thermal oscillations or instabilities (thermal runaway) in the
hydrogen burning shell as discussed by \citen{fih:90}, who emphasized
that the occurrence of this phenomenon is strongly related to the
choice of the time steps in the numerical computations. However, all
numerical experiments we performed have always provided negative
results.  Overall, there are no significant differences between
canonical and standard models along the RGB. This is a well-known
fact, since at the base of the RGB the surface convection reaches its
maximum extension and mixes back into the convective envelope
basically all the chemical elements previously diffused toward the
center.  The luminosity at the RGB tip ($L_{\rm tip}$) and the size of
the He core at the He-burning ignition ($M_{\rm cHe}$) shows only
minor differences: $M_{\rm cHe}= 0.497M_\odot$ and $\log(L_{\rm
tip}/L_\odot)=2.357$ for the canonical model, and $M_{\rm cHe}=
0.498M_\odot$ and $\log(L_{\rm tip}/L_\odot)=2.361$ for the
standard one. The values of $M_{\rm
cHe}$ and $\log(L_{\rm tip}/L_\odot)$ appear in fair agreement with
the values given by \citen{cc:93} in spite of the different
assumptions about the primordial metallicity. Some significant
differences, however, exist in comparison with the data provided by
\citen{fih:90} and \citen{fsih:95}: $\Delta{M_{\rm cHe}}\approx-0.015M_\odot$ -
$\Delta\log(L_{\rm tip}/L_\odot)\approx-0.06$ in the sense that present models
have a fainter luminosity and a smaller He core. However, these differences
might be understood in terms of changes in the adopted physical
scenario.

To close this section, we briefly comment on the helium flash, which
terminates the RGB evolution. \citen{fih:90} and \citen{hif:90} have
strongly claimed that during the He flash in a $Z=0$ model the outer
edge of the convective shell formed in the helium zone can extend into
H-rich layers, thereby mixing hydrogen back into the helium core. This
could produce relevant changes in the surface chemical abundance of
the star. In our present computations (but see also the extended
survey made by \cite{cc:93}) we did not find any evidence for this
phenomenon, which depends strongly on the competition between
convective and nuclear timescale. The former one is not known in the
simple mixing-length convection theory, but can only be
estimated. Therefore, no firm conclusions about the occurrence of
mixing events during the helium flash can be drawn without a more
extensive investigation.

\clearpage

\section{The non-standard evolutionary models: the effect of external
pollution}

\subsection{Evolutionary properties}

The absence of metal-free and the relative paucity of extremely
metal-poor stars has been sometimes explained by taking into account
the hypothesis of enrichment of the surface metallicity due to
accretion of metal-rich material through encounters with interstellar
gas clouds. Even if this hypothesis seems to be the most promising one
in order to explain the observational evidences, as far as we know it has
not been fully investigated until now. The evolutionary consequences
and observational implications of such a process have been estimated by
\citen{yos:81} and \citen{fsih:95}.

In this respect, the work by \citen{yos:81} is quite important as it
provides some reasonable estimates of the amount of accreted matter
and its heavy elements abundance. The amount of material accreted on a
star, due to encounters with gas clouds during its travel through the
Galaxy, depends on several parameters as, for instance, the
relative velocity between the star and the gas cloud, the cloud
parameters and so on. Making some realistic
assumptions about the value of these different quantities and the
parameters of the stellar orbits, \citen{yos:81} has estimated that the global amount
of material accreted on a star with mass of the order of $0.8M_\odot$
in a timespan of the order of $10^{10}$ yr has to be in the range
$(10^{-3} - 10^{-2})M_\odot$. As far as the metal
enrichment is concerned, it strongly depends on the details of the chemical
evolution of the Galactic matter. \citen{yos:81} has estimated that
after $10^{10}$ yr, due to external pollution, the surface metallicity
of an extremely metal-deficient star should be in the range
$0.0006\le{Z}\le0.01$ (depending on the stellar orbit properties). 
The effect on the stellar structures due to atomic diffusion or
convective mixing (effectively dilution), however, were taken into account
by means of simple considerations based on the known physical
properties of standard stellar models for very metal-poor
objects. 
Also in the paper by \citen{fsih:95} the accretion of metals
was not treated in a self-consistent way.

For instance, both analyses do not take into account the
possible changes in the thermodynamical properties of the envelope due
to the change in the opacity of the stellar matter. In addition, if
one is interested in investigating in detail the possible changes of
the evolutionary behavior of Pop~III stars due to
accretion of metal-rich matter onto the stellar surface, it is
necessary to verify if and when the atomic diffusion is able to bring
CNO elements below the edge of the convective envelope and into the
H-burning region, which can either be the core or, in later phases, a
shell. To test this last point,
we have decided to use the most simple accretion model
possible. We assume that instantaneous accretion of all the metal-rich
matter has happened immediately before the star reached the ZAMS. This
means that we simply modify the chemical composition of a certain
fraction of the external stellar layers before starting the
MS computations. This approach has the benefit to maximize the
"efficiency" of the combination of both processes (accretion + diffusion).

\begin{figure}[ht]
\centerline{\epsfxsize=0.9\hsize \epsfbox[0 194 624 500]{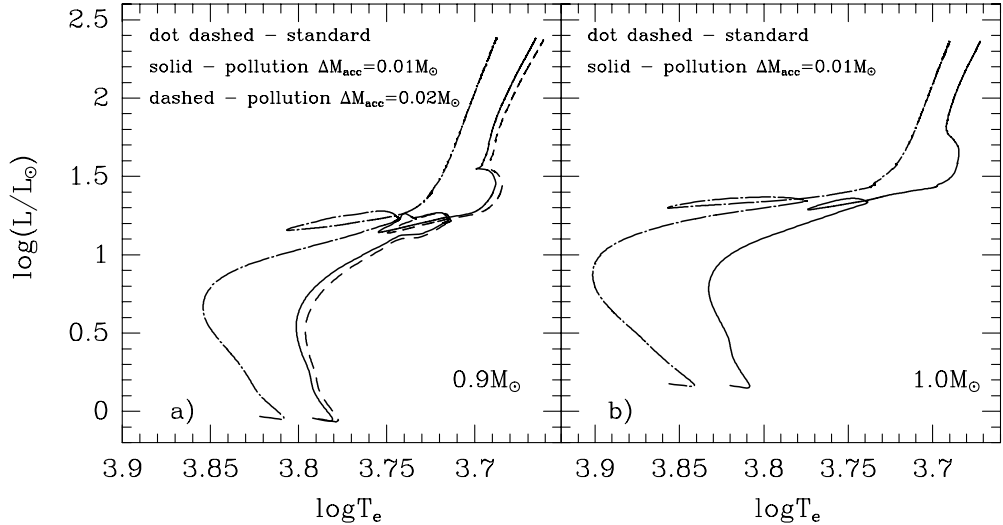}}
\caption{(a) The evolutionary paths in the HRD for a $0.9M_\odot$ 
stellar structure, computed under different assumptions about the amount of accreted metal
rich matter. (b) as (a) but for the $1M_\odot$ model.}
\protect\label{f:hrpol}
\end{figure}

In the experiments carried out, we have modified the chemical composition
of the outermost $0.01M_\odot$ of a model, by adopting as the chemical
composition of the ``accreted'' matter $Z=0.01$ and $Y=0.25$. In the
following, we are going to describe the resulting evolution of the
$0.9M_\odot$ and the $1M_\odot$ models. In Fig.~\ref{f:hrpol},  
we have plotted the evolutionary tracks in the HRD  and, for
comparison, repeated the evolutionary track of the standard
metal-free model with atomic diffusion (see \S~3). 

\begin{figure}[ht]
\centerline{\epsfxsize=0.7\hsize \epsffile{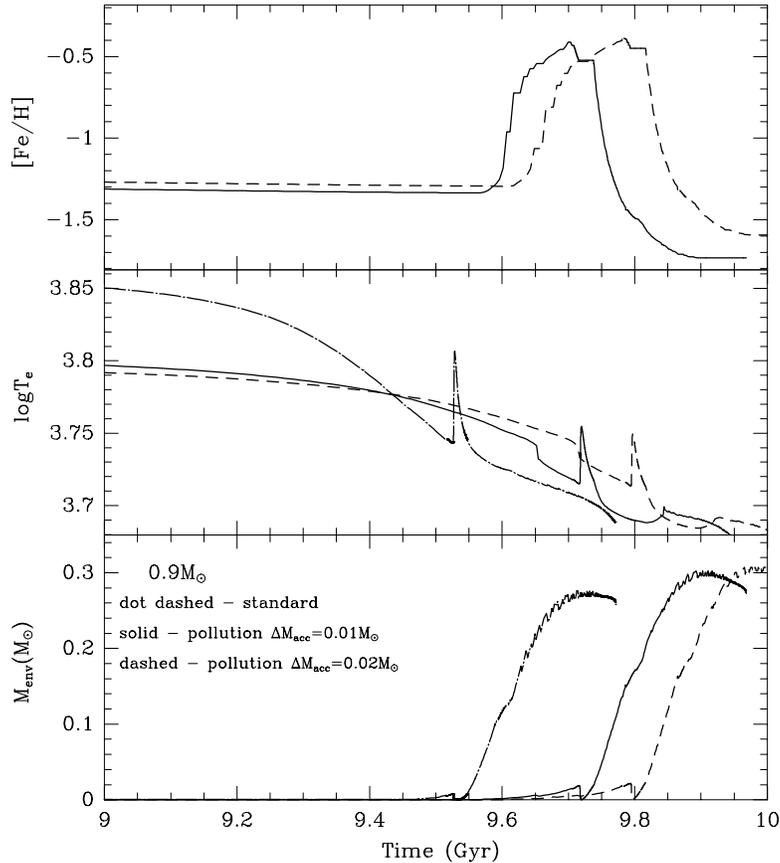}}
\caption{Evolution of the surface abundance of iron, the
effective temperature, and the size of the convective envelope for a $0.9M_\odot$
model computed taking into account different amounts of accreted
metal-rich matter.}
\protect\label{f:fetemc}
\end{figure}

The main result is that the accreted carbon never reaches the nuclear
burning regions. Some interesting features can easily be recognized in the
HRD, i.e., the shift toward lower effective temperatures of the
models with external pollution.  This is due to the
increase of the opacities in part of the stellar structure as a
consequence of the metals accreted. To be more specific, the
lower boundary of the polluted region of the star is located at
temperatures higher than $10^{6}$ K, well below the edge of the
convective envelope. It is also worth noticing that the occurence of
the CNO-flash along the SGB -- an event originated in the deep interior
-- is not affected at all by the accretion of metals.
This is indirect proof that atomic diffusion is not
able to bring CNO-elements into the nuclear burning region. In fact, had
it been the case, one would have witnessed an increase in the CNO-cycle
burning rate, eventually developing into a thermal runaway, during the
previous MS evolution.  The fact that during the central
hydrogen burning phase the evolutionary properties of the polluted
stars are not changed\footnote{One can easily notice that the chemical
pollution of the outermost layers has the effect of slightly
increasing the evolutionary lifetime (see Fig.~\ref{f:fetemc}). This
occurrence has to be 
related to the evidence that the polluted models are moderately
fainter than the standard ones.} significantly by the diffusion of
heavy elements -- i.e., that atomic diffusion is not efficient enough
to bring down the CNO-elements needed to start the CNO-cycle -- is
also confirmed by the behavior of the various energy sources all 
along the evolution. 

\begin{figure}[ht]
\centerline{\epsfxsize=0.7\hsize \epsffile{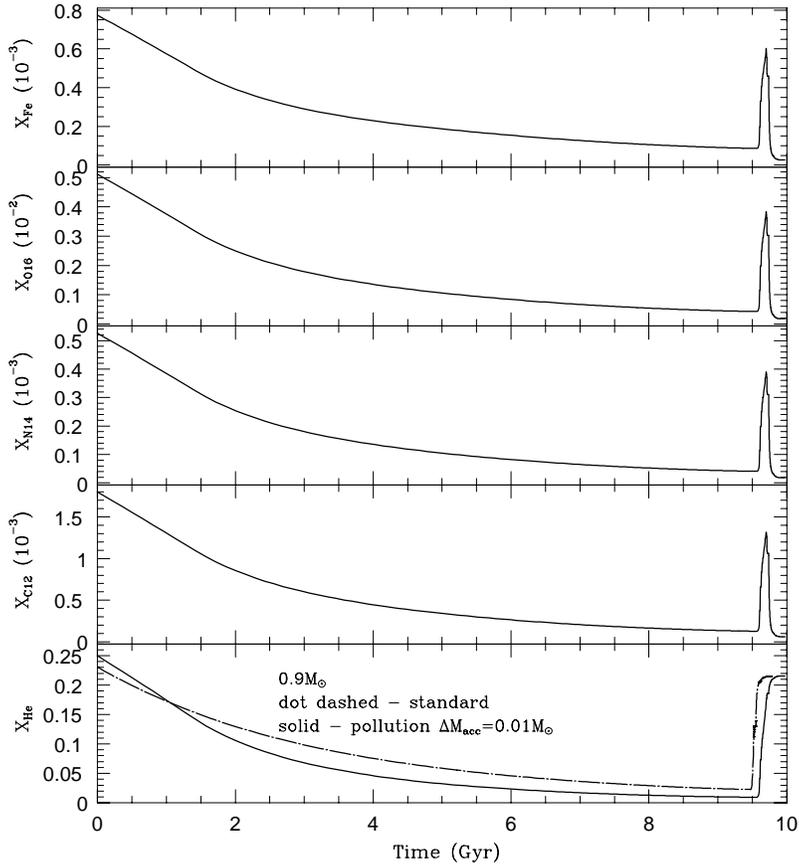}}
\caption{Evolution of the mass fractions of CNO-elements, iron and
helium, for a $0.9M_\odot$ model computed accounting for external
pollution. In the bottom panel, the run with time of the surface
helium abundance in the model including only diffusion has also been
plotted.}
\protect\label{f:chetime}
\end{figure}

During the SGB evolution, before the CNO-flash occurs along the way to
the RGB, a thin convective region appears in the envelope (see
Fig.~\ref{f:fetemc}, lower panel, at $t\approx 9.6$~Gyr in the case of
pollution of $0.01\,M_\odot$). Even if the thickness of this
region is very small ($\approx 0.02\,M_\odot$), basically all the
metals diffused during the MS evolution are dredged back to the
surface.  This is easily noticeable looking at the top panel of the
same figure (which displays the run of [Fe/H] with time) and also from
Fig.~\ref{f:chetime}. The increase in envelope metallicity is
accompanied by a decrease in effective temperature.  Then the
convective envelope goes even deeper, mixing therefore only matter
with original zero metallicity; this results in a slight decrease of
the surface abundances of the individual heavy-elements at $t\approx
9.7$~Gyr. The maximum abundances reached before this point are never
exactly the same as the initial one of the polluting matter, since the
diffused metals were already diluted with a metal-free environment.

Then the star experiences the thermal runaway; $T_{\rm eff}$ suddenly
increases and the convective envelope disappears, leaving the surface
abundances unchanged for a short while. When the runaway stops and the
track goes back to the normal evolution toward the RGB, envelope
convection sets in another time, this time reaching much deeper
($\approx 0.25\,M_\odot$).  At this point the surface metal abundances
decrease even more (more and more metal-free matter is mixed into the
convective envelope), while the stellar track moves toward lower
effective temperatures because of the progressively larger convective
region.  When the star is settling on its Hayashi track, the metal
abundance in the convective envelope is still decreasing; this
produces the "kink" which appears at the base of the RGB in
Fig.~\ref{f:hrpol}. It is due to the fact that the star is trying to
settle on the Hayashi track corresponding to the metallicity of its
convective envelope, but the metallicity is still changing due to the
deepening of the convective region; therefore the track has to move
toward larger $T_{\rm eff}$, since the RGB-location moves to larger
$T_{\rm eff}$ for decreasing surface metallicity.  Eventually this
process ends when the convective envelope has almost reached its
maximum depth, so that a small change in the extension of the
convective region does not appreciably change the surface metallicity,
and the star starts its standard RGB evolution.

Up to this point our calculations indicate that diffusion is not able 
to transport the accreted metals into the nuclear burning regions. 
The last possibility for this occurrence
takes place during the RGB evolution, where the
hydrogen-burning shell could be able to pass through regions previously reached
by the envelope convection at its maximum extension. This effect is
present in Pop~II stars and is the physical reason for the so-called
RGB-bump. Since the envelope metallicity, even if being very diluted
with respect to the beginning of the MS phase (by a factor of about 100), 
is significantly larger
than zero, a substantial amount of CNO-nuclei could be ingested into
the H-burning shell; this could cause a dramatic change in the
efficiency of the H-burning, of the same kind as the one experienced
on the SGB.  The outcome of our evolutionary computations rules out
this possibility, too, even in the very extreme case of our
assumptions about the pollution mechanism.  In all models we have
computed, the distance in mass between the inner point reached by the
convective envelope during its maximum penetration and the location of
the hydrogen-burning shell at the He-flash, is never less than
$\approx0.1M_\odot$ (we note, in passing, that the absence of
the RGB-bump constitutes another difference between Pop~III and
extreme Pop~II evolution).  Moreover, we recall that the evolutionary 
timescale along the RGB is too short for allowing atomic diffusion to bring down
CNO-elements from the point of deepest extent of the convective
envelope prior to the $3\alpha$ ignition in the helium core.

We have also checked if a larger amount of metal-rich accreted matter
could produce some significant change in the described evolutionary
properties of the models. To this end we have doubled the amount of
accreted matter ($M_{acc}=0.02M_\odot$), which is a factor $\approx 2$
larger than the maximum value suggested by \citen{yos:81}. One can
easily notice from the data plotted in Figs.~\ref{f:hrpol} and
\ref{f:fetemc} that the effect is very small and that the CNO elements
in the accreted matter never reach the H-burning regions.

Taking into account the fact that we have maximized the effects of
external pollution plus diffusion, we can safely conclude that the
accretion of a significant amount of metal-rich matter on the surface
of a metal-free star is not able to change significantly its
evolutionary behavior, except for a decrease of effective temperatures 
(due to the higher metallicity of the external polluted regions)
and a small increase in the MS lifetimes.  This is mainly due to the fact
that atomic diffusion is never efficient enough to bring CNO elements
into the inner stellar layers.

\subsection{Temporal evolution of the surface metal abundances in
polluted metal-deficient stars}

Since we have investigated the evolutionary effects of external
pollution plus atomic diffusion on metal-free stars by computing
self-consistent evolutionary models, we are able to show the run with
time of the surface abundances of the most relevant chemical
elements. This is a quite important issue since, indeed, it is potentially 
the only method available for discriminating between polluted Pop~III and
extreme Pop~II stars, when observing isolated field stars.

In Fig.~\ref{f:chetime} we have plotted the mass fraction of the
CNO-elements, helium and iron as a function of time all along the
evolution from the ZAMS until He ignition at the tip of the RGB. One
can easily notice that as a consequence of atomic diffusion the
surface abundances of all these elements are monotonically decreasing
during the MS phase.  However, as largely discussed in the previous
section, when the star approaches the RGB, the convective envelope
goes deeper inside the star, dredging up the elements which diffusion
has previously carried down in the structure. This occurrence has the
effect to produce the sharp increase of the displayed chemical
abundances; as soon as the outer convection continues to
deepen, it reaches the layers consisting of primordial matter,
mixes it with all the outer layers, and
the surface chemical abundances sharply decrease again.

The surface abundances evolution of a polluted Pop~III star is therefore
initially reflecting the pollution process, which we have concentrated
into a singular event at the beginning of the star's MS
evolution. Alternatively, continuous accretion or individual pollution
events at any time during the MS phase can be envisaged. The pollution
effect is modulated by that of diffusion, which will lead to
declining metal abundances similar to those as shown in
Fig.~\ref{f:chetime}. As soon as the envelope becomes convective --
which, in turn, depends on the surface metallicity as well -- the
previous diffusion history will be obliterated and only the mass
ratio between accreted material and the stellar convective envelope will
determine the observable abundances along the RGB.

\begin{figure}[ht]
\centerline{\epsfxsize=0.7\hsize \epsffile{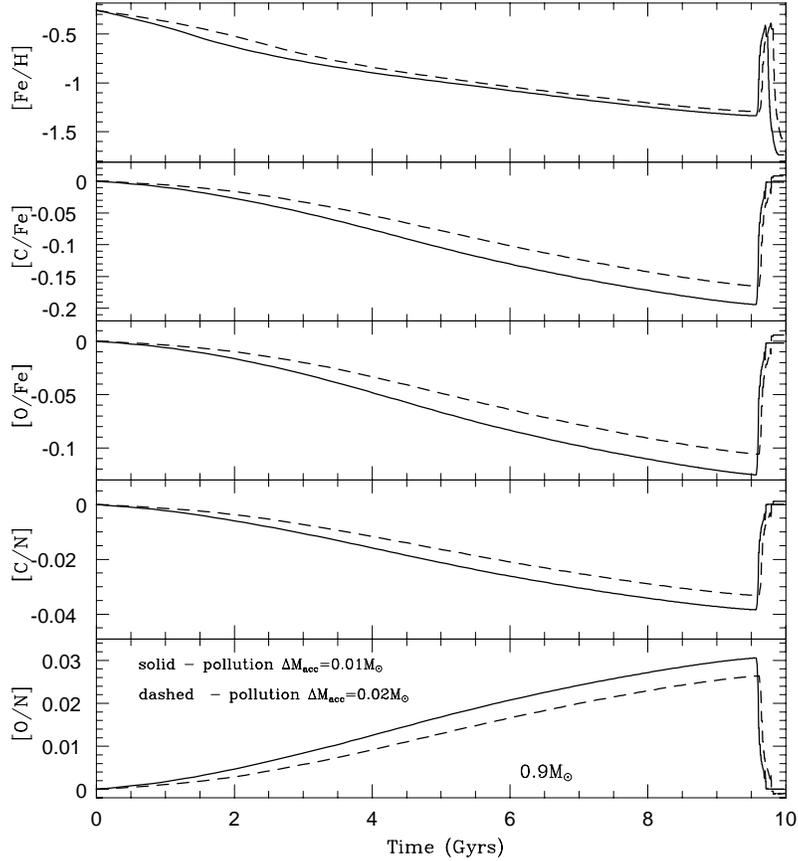}}
\caption{Abundance ratios as function of time between CNO
elements and iron, and between the CNO elements themselves, for a $0.9M_\odot$
model computed by adopting two different assumptions (as labelled) on
the amount of polluted matter.}  \protect\label{f:raptime}
\end{figure}

Since the abundance of the single various heavy elements cannot be obtained
directly by spectroscopic measurements, we plot in Fig.~\ref{f:raptime} 
the expected behavior with time of observable quantities, namely 
the abundance ratios [Fe/H], [C/Fe], [O/Fe], [C/N] and [O/N]. 

{From} the data plotted in this figure one expects that, when accounting
for atomic diffusion and external pollution, Pop~III stars should
show, during a significant portion of their evolution before reaching
the RGB, a slight overabundance of (O/N) and an underabundance of
(C/N), (O/Fe) and (C/Fe) in comparison with the Sun (we recall that a
scaled solar abundance has been assumed as the distribution of the
polluted matter), and more in general with respect to the same element
ratios in the polluting material.  However, one has to take into
account that the value of these under- and overabundances are quite
small (lower than the commonly adopted uncertainty of $\pm0.10-0.15$
dex in spectroscopical measurements) since the diffusion
velocities for C, N, O and Fe are quite similar and, moreover,
that in our experiments we have maximized the effect of pollution plus
diffusion. The element ratio showing the largest difference with
respect to its value in the polluting matter appears to be [C/Fe]
(again, of course, only during the MS phase).

Can the measurement of these abundance ratios in MS field UMPHS 
-- in particular [C/Fe] -- 
be of some help for discriminating between a very metal poor Pop~II MS star 
and a polluted Pop~III one?
Probably not very much, for two kind of reasons.
The first one is that, as repeatedly stressed before,  
in our experiments we have maximized the effect of pollution
plus diffusion. A steady slow accretion of metals, or 
an impulsive accretion by discrete amounts (which added together
should amount at most to about 0.01$M_{\odot}$) would have a smaller
effect on the surface abundance ratios, since diffusion has less time to work.
Moreover, diffusion should be effective also in Pop~II stars
and therefore a similar value for, 
i.e., the [C/Fe] ratio, would be observed at the surface of a true Pop~II star
or of a polluted Pop~III one, at least in the simplest hypothesis that  
the element ratios in the matter from which Pop~II stars originated and 
the matter accreted by Pop~III objects are basically the same.

Before concluding this section we want just to recall that,
as already emphasized previously, in our computations 
we have found no evidence for any deep mixing phenomenon at the He ignition, 
as suggested earlier by \citen{fih:90} and \citen{hif:90}.
In the hypothesis that this mechanism is efficient in the cores of Pop~III 
stars, the surface metal abundance (in particular C and N) would be enhanced
at the tip of the RGB, in contrast to the effect external pollution, which is
washed out during the RGB phase. According to \citen{fsih:95} a nitrogen rich carbon 
star with $-4.5 \le{\rm [Fe/H]}\le -2.0$ must be a PopIII stars.
Since the relevance of this issue for the problem of discriminating between
metal poor Pop~II and polluted Pop~III stars,
it is our intention to present a detailed investigation of the evolution 
through the He-flash in a forthcoming paper.

\section{Conclusions}

The main aim of this work was to investigate the possibility that a
low-mass Pop~III star could increase its original amount of CNO
elements in the core to a level which allows the ignition of the CN cycle,
thus modifying its evolutionary behavior from the one characteristic 
of metal-free object to the one typical of extremely metal-poor stars.

The only two channels allowing -- in principle -- for this occurrence,
are: the production of carbon by means of unconventional nuclear
reactions and the accretion of metal-rich matter through encounters
with molecular clouds. Both scenarios have been fully
explored by means of self-consistent evolutionary computations.

As for the possibility that non-standard nuclear reactions -- as the
ones suggested by \citen{mit:85} and \citen{wgg:89} -- could
significantly contribute, together with the canonical 3$\alpha$
reactions, to carbon production, it has been found that the most
important reaction is $^7{\rm Be}(\alpha,\gamma)^{11}{\rm C}$, which
dominates the carbon synthesis during the early MS phase. However, all
non-standard reactions are unable to increase the carbon abundance to
the level of $10^{-10}$, needed to ignite the CN cycle.  Therefore we
conclude that, at least in the explored mass range, it is necessary
only to account for the 3$\alpha$ reactions in the evolutionary models
in order to obtain a reliable estimation of the carbon production. The
uncertainty on the final abundance by mass of $^{12}{\rm C}$ or
$^{14}{\rm N}$, is of the order of a quite negligible 1\%. This
occurrence can be considered as plain evidence for the reliability of
current stellar models for metal-free objects.

The scenario in which a metal-free object accretes metal-rich matter
through encounters with molecular clouds, has been investigated by
computing evolutionary models accounting for both atomic diffusion and
chemical pollution, and by using reasonable estimates of the amount of
accreted matter.  The numerical computations have clearly shown that,
neither atomic diffusion during the MS phase nor outer convection
during the RGB evolution, are able to bring CNO-elements into the
nuclear burning region; in spite of the remarkable changes of the
heavy elements abundance of the outer layers, the evolutionary
behavior of a polluted Pop.~III star is always regulated by the
original chemical composition.

Since we have investigated only comparably well-known
particle-transport effects (convection and diffusion), one could
envisage that non-standard effects (for example, mixing induced by
rotation) help in transporting metals from the stellar surface to the
nuclear processing regions. Indeed, in globular cluster stars there is
strong evidence for such an additional mixing mechanism, whose
signature is evident by surface anomalies in CNO-elements as well as
Na and Mg (see \cite{kra:94} for a review). This mixing is simulated
in theoretical models (e.g.\ \cite{dw:96}) by an additional diffusion
process. The intention and result of these simulations is to transport
material from the hydrogen shell to the bottom of the convective
envelope. In our case, the direction would be opposite, but the mechanism
could be the same. While we cannot exclude such an additional effect
completely, there is a strong argument against its occurance in
metal-free stars: there are observational evidences (\cite{suntz:81};
\cite{gb:91}) supported by theoretical modelling and considerations
(\cite{sm:79}; \cite{cchar:95}) that the process starts only {\em
after} the red giant bump, i.e.\ when there is no molecular weight
gradient (barrier) between the outer shell and the rest of the
envelope. As demonstrated, true Pop~III stars, however, never reach
this phase because helium ignition sets in long before at very low
luminosities compared to those of Pop~II RGB-tips. This, in turn, is a
direct consequence of the hotter cores, the temperature of which is
determined by the shell temperatures. As we know, these are higher for
shells burning hydrogen via the $pp$-chains. We therefore consider it
unlikely that an additional mixing as in globular cluster red giants
would appear in Pop~III giants.

{From} the point of view of an 'external' observer it is extremely
difficult to discriminate between a polluted Pop~III field star and a
very metal poor Pop~II one; in spite of substantial differences in the
core physical conditions and energy production mechanisms, the
evolution in the HRD is qualitatively quite similar.  The effective
temperature of the star is basically regulated by the surface
metallicity, and the only feature peculiar of the Pop~III object is
the CNO flash along the SGB, a very fast and unobservable phase.  Also
the study of the surface abundance ratios does not appear to be of
very much help.  This means that, if the chemical pollution of the
stellar surface is effective, the still surviving Pop.~III stars could
be all disguised as extremely Pop.~II stars with no chance to
discriminate between ``true'' extremely metal-poor stars and polluted
Pop.~III objects.  The only possibility, as mentioned in the previous
section, is the occurrence of deep mixing phenomena at the He
ignition, which would produce a nitrogen rich metal poor carbon star;
this is a subject on which we will present a detailed investigation in
a forthcoming paper. Our models also predict a general trend for the
surface metal abundance of polluted Pop~III stars: it should steadily
decrease with evolutionary phase, with the exception of a brief
episode during the subgiant phase, when they are higher by a factor of
ten (Fig.~\ref{f:fetemc}). Given that enough UMPHS are observed and
that all of them had been polluted by single events during their
earliest evolution (\cite{shts:98}), this might be observable.

\newpage
\acknowledgments This work was supported by a DAAD/VIGONI grant. All
authors are grateful for the warm hospitality they received during
their visits at the institutes involved in the project. 
They also thank the organizers and participants of the 1999 MPA/ESO
conference on ``The First Stars'' for a very stimulating meeting.
Helpful discussions with V.~Castellani and P.~Marigo are 
acknowledged. 
F.-K.~Thielemann kindly provided his reaction library and added helpful 
discussions about hot pp-chains. This paper made use of the NASA ADS
system at its mirror sites at CDS, Strasbourg and ESO, Garching.

\end{document}